\documentclass[aps,prl]{revtex4}
\usepackage{graphicx}
\def\ee{\end{equation}}
\def\bea{\begin{eqnarray}}

\begin{document}

\title{Toy Models of Top Down Causation}

\author{Adrian \surname{Kent}}
\affiliation{Centre for Quantum Information and Foundations, DAMTP, Centre for
  Mathematical Sciences, University of Cambridge, Wilberforce Road,
  Cambridge, CB3 0WA, U.K.}
\affiliation{Perimeter Institute for Theoretical Physics, 31 Caroline Street North, Waterloo, ON N2L 2Y5, Canada.}
\email{A.P.A.Kent@damtp.cam.ac.uk} 

\date{September 2019; updated September 2020} 

\begin{abstract}

Models in which causation arises from higher level structures
as well as from microdynamics may be relevant to unifying  
quantum theory with classical physics or general relativity. 
They also give a way of defining a form of panprotopsychist property dualism, in
which consciousness and material physics causally affect one another. 
I describe  probabilistic toy models based on cellular automata that
illustrate possibilities and difficulties with
these ideas.
\end{abstract}
\maketitle
  
\section{Introduction}

The reductionist paradigm for theoretical physics suggests that 
the properties of complex structures, including their dynamics, can be understood as a 
consequence of those of their elementary components.
It is not easy to characterise precisely what this means in all cases.
Is a space-time or the vacuum state of a quantum field a complex
structure, for example?  And if so, what are their elementary
components?    Is a bare quark an elementary component or a 
mathematical fiction?   Is quantum entanglement a counter-example to 
reductionism or just an illustration that the concept needs to 
be framed more carefully?   

Nonetheless, it is widely agreed that some appropriately nuanced 
and qualified version of reductionism has been extremely successful, so much so
that many theorists seek unified theories in which
all of physics is characterised by some theory of the initial conditions
together with relatively simple (though seemingly probabilistic) dynamical laws.    

We should distinguish this strong but quite arguable stance
from the stronger and fairly indefensible one that understanding the 
fundamental laws is the only really important task in science.   
As Anderson compellingly argued in his classic essay \cite{anderson1972more}, solid-state physics, 
chemistry, biology, psychology and other higher-level theories 
produce new behaviours and new laws that require great inspiration
and creativity to find and understand.   But Anderson was nonetheless 
a card-carrying reductionist: 
\begin{quote}
{\it The reductionist hypothesis may still be a topic for controversy
among philosophers, but among the great majority of active
scientists I think it is accepted without question.   The workings
of our minds and bodies, and of all the animate or inanimate matter
of which we have any detailed knowledge, are assumed to be 
controlled by the same set of fundamental laws, which except
under certain extreme conditions we feel we know pretty well.}\cite{anderson1972more} 
\end{quote} 

Chalmers' \cite{Chalmers2006-CHASAW} distinction between types of emergence is very helpful 
here.    High-level phenomena are {\it weakly} emergent when they are unexpected, but deducible in
principle (even if not currently in practice) from fundamental
theories.    
They are {\it strongly} emergent if they are not deducible even in
principle. 
Reductionists aim for a relatively simple universal theory in which there are
no examples of strong emergence, but should be comfortable with weak
emergence. 
A representative survey would be very interesting,
but my guess is that Anderson's characterisation still holds true
today: most scientists believe 
we already know enough of the fundamental theory to understand non-extreme regimes,
and in particular would deny that the emergence of 
(quasi-)classicality from quantum theory, or (pace Chalmers
\cite{Chalmers2006-CHASAW, chalmers1996conscious})
of consciousness
from classical or quantum physics, are clear examples of
strong emergence.   On the other hand, these questions are hotly
debated among scientists working on consciousness and on 
quantum foundations.   

Although the boundaries of reductionism may be slightly fuzzy, we can
certainly produce models or theories that are clearly beyond them and unambiguously
anti-reductionist.  One example would be a theory that predicts 
qualitatively different dynamical equations for different 
types of molecule.\footnote{For the theory to be unambiguously
anti-reductionist, it should not be possible to derive these
equations from some simpler unifying principle.   
For example, dynamical collapse models \cite{ghirardi1986unified,ghirardi1990markov}
are not anti-reductionist, 
although in a sense they predict different behaviours 
for molecules of low and high mass: these predictions all follow
from the same stochastic differential equation.} 
Another would be a vitalist theory that
predicts that living creatures disobey some conservation
law of classical mechanics. 
The general consensus is that we should assign low priors
to such theories, not only because reductionism has been successful, 
but also because reductionist theories tend to be more elegant, and
aligning credence with elegance has also been a very successful
scientific methodology.       

There is, though, more serious interest in better motivated
models that do not fit the current mainstream reductionist paradigm.  
Consciousness, the topic of this special issue, {\it seems} 
to give one than one causal narrative -- mind and matter {\it seem} to affect
both themselves and each other.
Yet the causal effects of matter on matter also seem enough for a
complete description of the material world: there is no physical evidence that 
the known fundamental laws of physics don't suffice to describe 
the behaviour of the brain. 
There are (controversial) mainstream reductionist stances on this
(e.g. \cite{dennett1993consciousness}), but 
also well-known (also
controversial) arguments (e.g. \cite{jamesautomata,nagel1974like,chalmers1996conscious,penrose1990emperor,penrose1994shadows})
against the reducibility of consciousness to known physics. 
There has been an upsurge of interest lately in 
exploring alternative ideas involving new (or non-standard) 
physical hypotheses
(e.g. \cite{chalmersfqxi,hardy2015bell,hardy2017,tegmark2014consciousness,tegmark2014consciousness2,tegmark2016improved,kremnizer2015integrated,okon2016back,cmdraft}). 
Several of these have drawn inspiration and motivation
from work on ``integrated information theory'' (IIT)
\cite{oizumi2014phenomenology}
which, although open to many criticisms (e.g. \cite{barrett2014integration,
  cerullo2015problem, 
  barrett2019phi}, gives a mathematical
framework to explore and generalise as well as a connection to empirical
data.

Ideas of top-down causation have been mooted in the
context of quantum theory and more broadly (e.g. \cite{aharonov2018completely,
  ellis2018top}). 
The scope for top-down causal models of
consciousness has not been extensively explored, and 
even the meaning and scope of top-down causation is not fully elaborated. 
This paper aims to give one framework for discussion,
by describing some simple toy models, inspired by cellular automata,
which illustrate possible ways in which higher level structures 
act causally on the microscopic components, as well as vice versa.  
It should be stressed that these are not
meant to capture most realistic features of the world.
The aim is to illustrate scope for more realistic models that use
a similar mechanism to combine types of causation. 

\section{Cellular Automaton $110$}

Our toy models are based on cellular automata, but are not meant in 
the spirit of the well-known research
programmes aiming to describe nature at a fundamental level in terms of cellular
automata \cite{wolfram2002new,t2016cellular}.    We use cellular automata
simply as convenient illustrations.   

Wolfram \cite{wolfram1983statistical,wolfram2002new} classified the
$256$ elementary 
one-dimensional cellular automata.   These are defined by binary states 
with a time step rule in which the states of cell $n$ at time $t$
are determined by those of cells $n-1,n,n+1$ at time $(t-1)$. 
He noted the particularly interesting and complex behaviour of the automaton
defined by rule $110$ in his classification.  
Again, we pick this out not out of any fundamental preference --
higher-dimensional cellular automata such the Game of Life
\cite{gardner1970mathematical} could equally well be used,
for example -- but for simplicity of illustration.   

Rule $110$ is defined by 

\begin{figure}[h]
\centering
\includegraphics[width=0.5\linewidth]{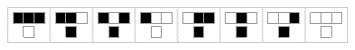} 
\caption{The rule 110 cellular automaton.   The states of cells
$n-1,n,n+1$ at time $(t-1)$, given on the first row, determine that
of cell $n$ at time $t$, given on the second row. }
\label{110def}
\end{figure}
Wolfram had previously suggested \cite{wolfram1986theory} that rule $110$ is Turing complete,
a result subsequently proved by Cook \cite{cook2004universality}; another discussion
of the result is given in Ref. \cite{wolfram2002new}. 
We review here some of its known properties, using results and images 
generated by Wolfram Mathematica and resources \cite{workbook}
from the Wolfram Data Repository, which helpfully includes routines that
reproduce many interesting diagrams originally given in Ref. \cite{wolfram2002new}. 

The rule generates a regular repeating array with period $14$, 
known as the ``ether'':
\begin{figure}
\centering
\includegraphics[width=0.5\linewidth]{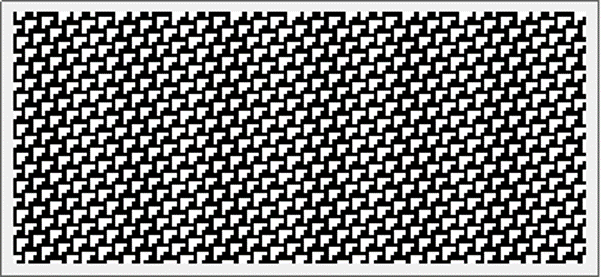} 
\caption{The ether.}
\label{ether}
\end{figure}
We will take this to be the analogue of a ``vacuum'' or ``background
state'' in the toy models defined below.

Some finite dislocations in the lattice-like 1D structure of a row of the
ether can propagate regularly.   These so-called ``gliders'' generate quasi-particle-like tracks 
in the ether.   Cook  \cite{cook2004universality} classified a variety of rule $110$ gliders, including some
infinite classes:
\begin{figure}
\centering
\includegraphics[width=0.5\linewidth]{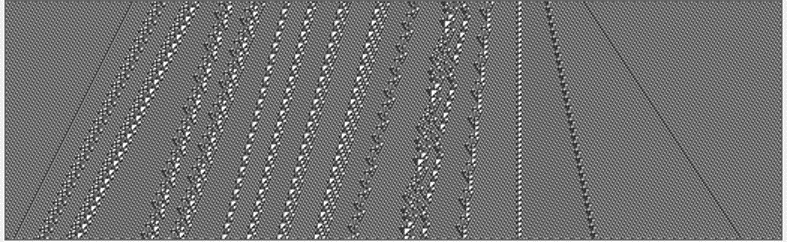} 
\caption{Gliders.}
\label{gliders}
\end{figure}  
Colliding gliders undergo ``interactions'' that are superficially
\footnote{Of course, these are deterministic classical interactions,
not complex quantum interaction amplitudes.}
reminiscent of Feynman
diagrams, typically producing a finite number of stable new gliders
after a finite interaction time:
\begin{figure}
\centering
\includegraphics[width=0.27\linewidth]{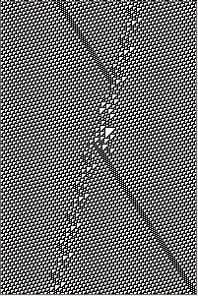}\hspace*{4mm}
\includegraphics[width=0.27\linewidth]{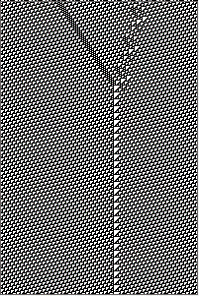}\hspace*{4mm}
\includegraphics[width=0.27\linewidth]{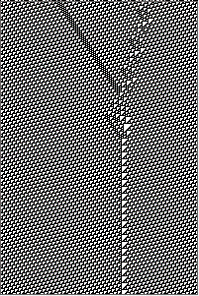}\hspace*{4mm}
\caption{Interactions.}
\label{interactions}
\end{figure}
(For more discussion of the general phenomena of glider ``particles''
and background ``domains'' in cellular automata, see e.g.
Refs. \cite{crutchfield1989inferring,hanson1992attractor,crutchfield1993turbulent,crutchfield1995evolution}.) 

We can define a very simple model of errors or noise in these
structures by considering the possibility of a single bit flip
on the first row.   One might motivate this by supposing that 
there is something particular about the system at $t=0$ that
makes errors or noise possible on the first row, and only that row,\footnote{Or so much more likely that
we can neglect the possibility of errors on other rows.} with
error probability low enough that we can neglect the possibility
of two or more errors arising 

\begin{figure}
\centering
\includegraphics[width=0.27\linewidth]{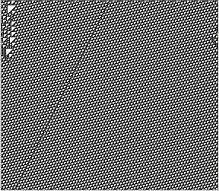}\hspace*{4mm}
\includegraphics[width=0.27\linewidth]{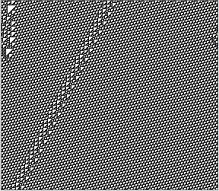}\hspace*{4mm}
\includegraphics[width=0.27\linewidth]{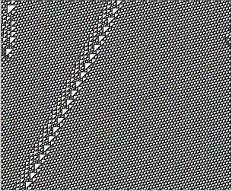}\hspace*{4mm}
\caption{Perturbing the ether at a point distant from a glider.}
\label{ether.perturb}
\end{figure}

If we consider a glider propagating in the ether, with a single 
bit of the initial state flipped at a site in the ether that is
far from the glider, the effect tends to be simply to cause some
ripples in the ether that propagate for some time and then peter 
out without interacting with the glider.   The glider's propagation
is thus unaffected, as Fig. \ref{ether.perturb} illustrates. 

\begin{figure}
\centering
\includegraphics[width=0.27\linewidth]{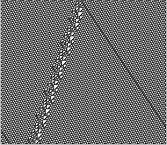}\hspace*{4mm}
\includegraphics[width=0.27\linewidth]{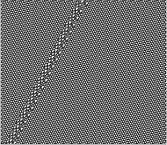}\hspace*{4mm}
\includegraphics[width=0.27\linewidth]{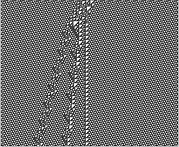}\hspace*{4mm}
\caption{Perturbing the ether at a point close to a glider.}
\label{ethercloseperturb}
\end{figure}

However, if we flip a bit close to a glider, it can interact in a way 
that permanently alters 
the number and type of gliders.   Fig. \ref{ethercloseperturb} shows the same glider states
with one initial bit flipped.      Only the second is asymptotically unaffected.     

Perturbations affect interacting gliders similarly.  A perturbation
distant from interacting gliders generally peters out without
affecting them.   However, perturbations in the vicinity of one or
more interacting gliders may alter the types and/or number of gliders 
in the final or asymptotic state.    

\begin{figure}
\centering
\includegraphics[width=0.9\linewidth]{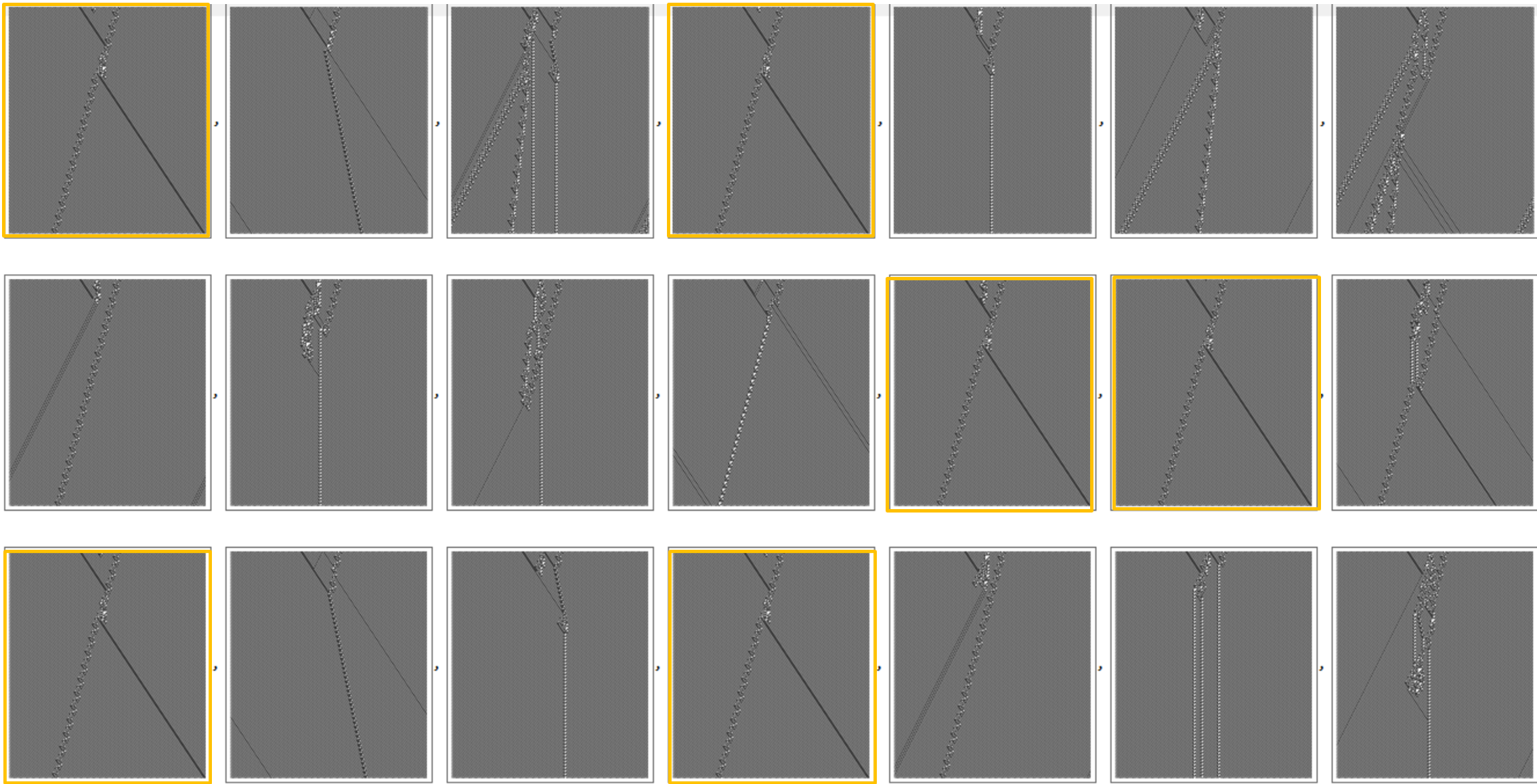}\hspace*{4mm}
\caption{Perturbations near interacting gliders.   The highlighted
  examples leave the final states asymptotically unchanged.}
\label{ifpu}
\end{figure}

\begin{figure}
\centering
\includegraphics[width=0.9\linewidth]{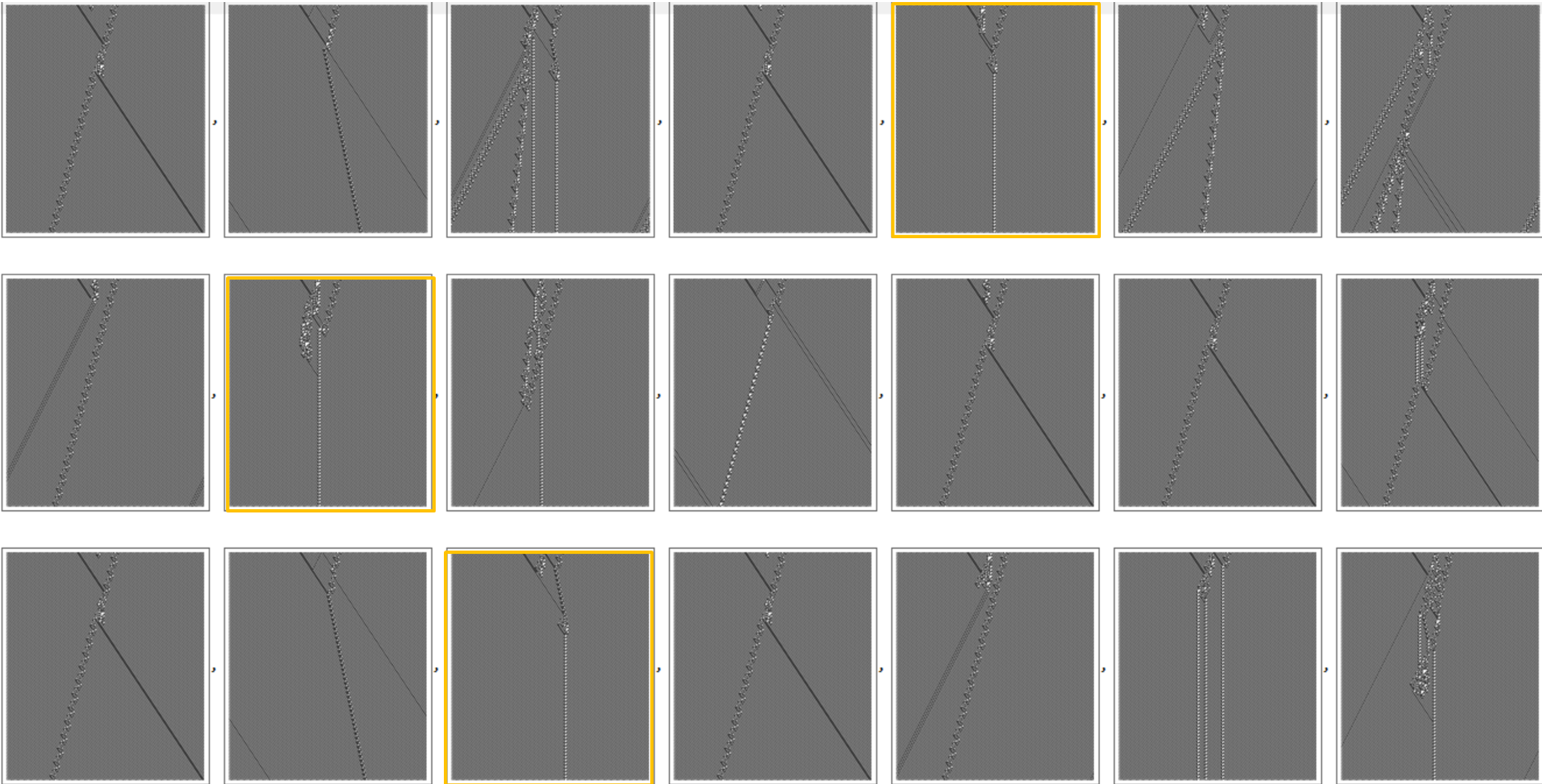}\hspace*{4mm}
\caption{Perturbations near interacting gliders.   The highlighted
  examples produce the same asymptotic final state, a single glider.}
\label{ipsf}
\end{figure}

Fig. \ref{ifpu} shows a pair of interacting gliders with a 
single flipped bit, whose site runs sequentially through $21$   
sites initially located between the gliders.   
Of these perturbations, the $1$st, $4$th, $12$th, $13$th, $15$th
and $18$th leave the final glider states, highlighted, are asymptotically unchanged.
The $5$th, $9$th and $17$th, highlighted in Fig. \ref{ipsf},
all produce the same new asymptotic
final state, consisting of a single glider.   

\section{Probabilistic models based on cellular automata and top-down causation}

\subsection{A simple probabilistic model}

We can formalise the model above as a $1D$ cellular automaton
whose states are defined on sites labelled by the integers,
at times also labelled by the integers.   
When the error probability $p$ is zero, it is a deterministic 
type $110$ automaton.   The ether, or the ether with a
single glider, then propagate indefinitely without
perturbation.   A pair of gliders may approach one another
from infinity, interact, and produce some number of 
outgoing gliders, which then propagate indefinitely. 

We may also take the model to have finitely many spatial
sites, with periodic boundary conditions.   In this case,
with appropriate numbers of sites, the ether and a single
glider state may still propagate indefinitely.    
If a pair of gliders with different velocities are evident during some time
interval, they will, so to speak, interact in both the past and
future.   If the interaction products contain two or more gliders
with different velocities, they in turn will interact, and the
asymptotic behaviour may be quite complex. 
We can avoid this by defining the model only for a finite time interval, short
compared to $N/v$, where $N$ is the number of sites and $v$ the 
maximum glider speed.      

We suppose that there is some probability $p>0$ of an error occurring
on the row of sites at $t=0$.   An error flips the bit value of 
a single site, so that it takes the opposite value to that predicted
by the deterministic dynamics from the state at $t=-1$.  
To simplify, we suppose that there is no probability of more than
one error, and that errors are restricted to sites $x$, where
$-M \leq x \leq M$, where $(2M+1) \leq N$ if there are finitely many
($N$) sites.   The errors in this region have uniform probability,
so that each site in the region has error probability
$\frac{p}{2M+1}$. 

The discussion of the previous section then applies: errors
sufficiently far from any gliders at $t=0$ will typically
peter out before interacting and have no effect on the final
or asymptotic late time glider states;
errors close to gliders can alter the number and type 
of final or asymptotic late time gliders.     

\subsection{Incorporating top down causation}

We now consider modifying the dynamics by assigning probability
weight factors to final glider states conditional on initial glider
states.   

One simple rule is to assign probability weight factor 
$1$ to final states that are the same as the initial state for single glider propagation and
two glider interactions, and weight factor $0$ to distinct states. 
Formally, 
\begin{eqnarray}
p_{\rm mod} ( G_f | G_i ) &=& C w (G_f | G_i ) p (G_f | G_i ) \, , \\
p_{\rm mod} ( G^1_f , G^2_f , \ldots , G^n_f | G_i ) &=& 0
\qquad{\rm~for~} n \neq 1 \, , \\
p_{\rm mod} ( G^1_f , G^2_f | G^1_i , G^2_i ) &=& C' w (G^1_f , G^2_f | G^1_i , G^2_i )
p (G^1_f , G^2_f | G^1_i , G^2_i ) \, , \\
p_{\rm mod} ( G^1_f , G^2_f , \ldots , G^n_f | G^1_i , G^2_i ) &=& 0
\qquad{\rm~for~} n \neq 2 \, . 
\end{eqnarray}
Here 
\begin{equation}
w (G_f | G_i ) = \delta_{G_f , G_i } \, .
\end{equation}
Multiple glider states are listed from left to right and so
\begin{equation}
 w (G^1_f , G^2_f | G^1_i , G^2_i ) = \delta_{G^1_f , G^2_i} \delta_{G^2_f  ,
   G^1_i }  \, 
\end{equation}
for colliding gliders, while
\begin{equation}
 w (G^1_f , G^2_f | G^1_i , G^2_i ) = \delta_{G^1_f , G^1_i} \delta_{G^2_f  ,
   G^2_i }  \, 
\end{equation}
for gliders that never collide. 
The expression $p(G_f | G_i )$ is the probability of the final state containing
(only) the single glider $G_f$ when the initial state contains glider $G_i$ in the model of the
last subsection; $p (G^1_f , G^2_f | G^1_i , G^2_i )$ is the probability
of the final state containing (precisely) the pair $G^1_f ,G^2_f$ when the
initial state contains the pair $G^1_i , G^2_i$; $C$ and $C'$ are rescaling factors chosen so that the
probabilities of all possible final states sum to $1$ for a given
initial state.   

This rule is understood as applying to the system as a whole.
It does not alter the deterministic dynamics of rule $110$, 
and so its effect is to alter the probabilities of errors in the 
state at $t=0$, which are the only probabilistic feature of the
toy model.   
For example, for an initial state containing a single glider $G$, 
it slightly increases the probability of errors at sites
(such as those far from the glider) where they do not affect
the glider propagation, increases the probability of no error,
and eliminates the possibility of errors occurring at sites
where they would alter the asymptotic glider propagation.   
It has similar effects for initial states containing two
gliders $G$ and $G'$. 
Effectively, the rule acts to suppress errors in glider
propagation, ensuring the stability of one and two glider
states, which would otherwise be menaced by possible errors in the 
microdynamics.

A variation is to assign probability weight $1$ to specified 
final state outcomes of two glider interactions, and $0$ otherwise,
while retaining the weights above for single glider states
Thus
\begin{equation}
w (G_f | G_i ) = \delta_{G_f , G_i } 
\end{equation}
as above but $w ( G^1_f , G^2_f , \ldots , G^n_f  | G^1_i , G^2_i )$ may have a more general
form.
For example, we might take $w (G_f | G^1_i , G^2_i ) = 1 $ for some
specified final state $G_f$, and zero for all other final states.
This ensures that
initial glider states $G^1_i ,G^2_i$ always produce final state $G_f$.   
however small the unmodified probability of this outcome is,
so long as it is nonzero.  

\subsection{Compatibility with standard temporal and Minkowski causation} 

Framed as above, these modified toy models may appear to
involve something like instantaneous action at a distance,
since the probability of error at a given site at $t=0$ effectively
depends on the type and number of gliders at distant sites 
at the same time.  
If we think of the models as capturing the behaviour of
particles (modelled by gliders) propagating in a background
(the ether) with stochastic fluctuations (the errors), 
in some non-relativistic limit of a theory in relativistic space-time,
this may seem to involve retrocausation: the probability of an error at a site
depends on the final glider states in regions in its causal future. 

While the models certainly could represent features of 
theories with non-standard causation, they are compatible with
standard causation, even for relativistic theories.    
We can take the relevant glider speeds to be below light speed
in such theories.   The gliders contained in the state at $t=0$ depend
deterministically on those contained in the states at $t<0$.  
We can thus equally well understand the probability of error
of any site at $t=0$ as determined by the glider states
at suitably large negative $t$, when the gliders are within the site's
past light cone.    
Interpreted in this way, errors at $t=0$ are causally determined by
glider states at large negative $t$, according to laws that ensure
specific glider states at large positive $t$.   
For example, one might imagine the models as capturing essential features
of some deeper theory in which this causal determination is 
made more explicit, by degrees of freedom that
carry information away non-superluminally from negative time glider
states to sites throughout the ether and influence the
error probabilities at $t=0$ appropriately.   
 
\section{Discussion}

\subsection{Panprotopsychist models of consciousness}

There are reasons to consider physical models of consciousness
that feature top down causation (although, as with every approach
to consciousness, there are also problems and counterarguments).
One line of argument runs as follows.   There 
are evidently physical correlates of consciousness, namely human
brains.   If there is a fundamental physical law associating conscious
states to physical systems, it seems unlikely that it associates
consciousness to brains and to nothing simpler:   
brains seem too complex as physical systems to be the elementary
referents of such a law. 
Full-blown panpsychism, in which every elementary particle has 
an associated elementary consciousness, is a possible alternative, 
but comes with many problems \cite{seager1995consciousness,chalmers2017combination} 
and does not seem to fit naturally with
neuroscientific data and our conscious self-observations or those
reported by others.    
The intermediate option of panprotopsychism \cite{chalmers2015panpsychism}, according to which
elementary consciousness is associated with some
physical systems (whose nature remains to be specified) larger than elementary particles and smaller than
brains, shares some of the problems of panpsychism, but allows more
possibilities that might fit with empirical observation.
Taking panprotopsychism seriously means accepting {\it some} sort of new
physical law(s) associating the relevant systems with consciousness.

Our probabilistic models based on cellular automata can be taken as toy
models of interactionist panprotopsychism. 
In these toy models, the elementary bits at each site
are meant to correspond to elementary components, and
the deterministic dynamics of rule $110$ and unmodified error probability
rules correspond to the elementary microdynamics.    The gliders represent
physical systems associated with elements of consciousness, which
we might take to be quales or (if we stretch the present models even more
unrealistically in order to illustrate how the idea might be extended) thoughts that 
we can represent by a sentence such as ``I see blue''.   
The first modified versions of the model, in which the error probabilities
are redefined to ensure that single gliders and pairs of gliders
propagate unaffected by errors, then correspond to toy models in
which panprotopsychist consciousness ensures error suppression 
at the level of consciousness, in the sense that quales (or thoughts)
propagate unaffected in the substrate, despite errors in
the microdynamics.   
The second modified versions redefine the error probabilities to ensure that pairs of gliders produce specified outcomes
that would not arise in the absence of errors.   These correspond to toy
models in which panprotopsychist consciousness is equipped with its
own dynamics, which overrides the dynamics of the substrate when 
the two conflict.     

An argument in favour of something like this picture is that, if
particular physical structures are indeed singled out as having an
elementary proto-consciousness by fundamental physical laws, it is
arguably natural that these physical structures should also feature in the fundamental
dynamical laws.  One might even speculate that nature has
probabilistic laws because of the need to combine dual causalities, of
matter and mind. 

It is helpful to compare the pros and cons of this line of thought
with those of a similarly panprotopsychist form of epiphenomenalism.
This would associate consciousness in a lawlike 
way to specified physical structures, without modifying the
microdynamics.   
The problem with this and other types of epiphenomenalism, as William
James first pointed out \cite{jamesautomata}, 
is that they leave all the apparently evolutionarily adaptive properties of consciousness
unexplained.   If physical laws of consciousness do not affect the
microdynamics, then we and other creatures would function equally well
if we were unconscious zombies, or if pleasure and pain were
uncorrelated with evolutionary advantage, or if our consciousnesses
were focussed on information that had no relevance to our survival
or well-being, or if we had ``locked-in'' consciousnesses disconnected
from any of our communications.  
On this view, we have to accept that not only the existence of
consciousness, but the apparent fine-tuning of its specific features,
are just astonishingly convenient coincidences.     

In contrast, there is scope for more convincing explanations
of the evolution of consciousness and of some of its features if dynamics give it a genuinely
causal role in behaviour.    It seems plausible, for example, 
that effectively coupling two types of dynamical rule 
allows evolution to more easily produce stimulus-response circuits that are more
stable or follow higher-level reasoning.  
It also seems plausible that evolution would use this coupling to
allow creatures to communicate their conscious states to one another.
This would allow them to coordinate their behaviour better than 
communications that are influenced only by the microdynamics of their
physical substrates, since in these models their behaviour may be
directly affected by their conscious states.    
Models in which consciousness acts causally, via laws involving its
complex physical correlates, also seem to offer some scope
for explaining the correlation of pleasure (pain) with evolutionary (dis)advantage.
A pain is something a conscious mind attempts to avoid, arguably by
definition, and if the dynamics of conscious states reflect this, 
then evolution could naturally exploit this dynamics if
disadvantageous physical situations created physical (brain)
states that involved subsystems associated (via the hypothesized
laws) with avoidant conscious states.   

Even if these arguments can be made convincing, it would still seem a surprising and fortunate coincidence that, somewhere
in the evolutionary chain, and perhaps very early, life took a
material form that had proto-consciousness, and that matter and
consciousness were associated in such a way that evolution was able
to make use of the dynamical rules that give consciousness causal
effect (via its material correlates) on matter.  
A priori, one might imagine that, if there are simple laws of
psychophysical parallelism and simple associated dynamical laws,
they need have nothing to do with self-replicating chemicals 
or organic information processing systems.   
So it is fair to ask how much fine-tuning interactionist
panprotopsychist theories could explain,
and how much they would still leave unexplained.  
Still, a partial explanation is better than none, and we also
need to be clear whether we could possibly hope for a fuller 
explanation given our present conceptual frameworks. 
After all, we are conscious. 
Anyone who finds conceivability arguments \cite{iepconscious,
  chalmers1996conscious} persuasive has to 
accept this, and all the features of our consciousnesses, as marvellous yet 
contingent features of our universe.  
On this view, we might hope that relatively simple laws characterise our
consciousness and explain its evolution, but we can't
hope for an argument that the laws must take the form they do. 

Closer analysis of all these arguments would undoubtedly be valuable. 
It would also be interesting to develop more sophisticated toy models, in which we can see
rudimentary creatures evolving in a simple environment 
via modified dynamics.

\subsection{Quantum theory, gravity and classical physics}

As these toy models illustrate, probabilistic theories of microdynamics can be simply modified so
that structures at two or more levels play roles in 
the fundamental laws.   This makes it easy to build and 
explore models with top down causation.   
Such models could also potentially be relevant to unifying quantum theory and gravity.
For example, one could imagine space-time emerging from a
fundamentally quantum theory, within a theory in which it is equipped with its own
independent dynamical laws; in such a theory, both space-time and  
its quantum constituents would causally affect one another, with neither reducible to 
the other.      
The same type of relationship is possible between classical and
quantum degrees of freedom within a fixed background space-time.
Classical physics is normally thought of as emerging from and 
reducible to quantum theory (by Everettians; see e.g. \cite{saunders2010many}) or
some extension of quantum theory that does not radically alter the
dynamics (by non-Everettians who believe some
extension is needed to resolve the measurement problem). 
The latter looks plausible (e.g. \cite{kent2015lorentzian,kent2017quantum}) and the simplest 
possibility, but it is interesting to ask how strongly
empirical evidence constrains more general theories
\cite{kent1998beyond,kent2013beable,kent2017quantum,kent2020hodology} 
that support this type of dual causation. 

\section{Acknowledgements}
I am very grateful to the organisers and participants of 
the Oxford 2019 ``Models of Consciousness'' conference,
at which this work was presented; many of their 
comments and criticisms were very helpful. 
I would also like to thank anonymous referees for constructive
criticisms and suggestions.    
This work was supported by an FQXi grant, by UK Quantum
Communications Hub grant no. EP/T001011/1 and by Perimeter Institute
for Theoretical Physics. 
Research at Perimeter Institute is supported
by the Government of Canada through Industry Canada and by the
Province of Ontario through the Ministry of Research and Innovation.

\section*{References}

\bibliographystyle{unsrtnat}
\bibliography{qualia2}{}
\end{document}